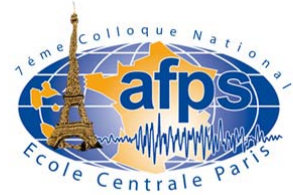

# Comparaison entre calculs de vulnérabilité sismique et propriétés dynamiques mesurées

**Clotaire Michel\* — Philippe Guéguen\*,\*\*— Pierre-Yves Bard\*,\*\***

*\* Laboratoire de Géophysique Interne et Tectonophysique (LGIT)*

*CNRS/Université Joseph Fourier*

*1381 rue de la Piscine, 38041 Grenoble cedex 9*

*cmichel@obs.ujf-grenoble.fr*

*\*\* Laboratoire Central des Ponts-et-Chaussées*

*58 boulevard Lefebvre, 75732 Paris cedex 15*

RÉSUMÉ. *Les méthodes d'analyse de la vulnérabilité sismique à grande échelle utilisent des formules et des courbes calculées à partir d'hypothèses simplificatrices pour lesquelles les incertitudes sont rarement précisées. Il est encore plus rare de les voir confrontées à des données expérimentales in situ. Nous avons donc enregistré les vibrations ambiantes et déterminé les caractéristiques modales (fréquences propres, déformées modales et amortissement) de 60 bâtiments grenoblois de types variés (béton et maçonnerie).*

*La connaissance des fréquences de vibration dans le domaine linéaire est capitale pour le dimensionnement dans les règles parasismiques. Nous avons donc comparé les formules donnant la période propre utilisée dans les codes à ces données expérimentales. Outre une variabilité sous-estimée, il apparaît que seule la hauteur du bâtiment apporte statistiquement de l'information sur sa fréquence propre, des formules simples avec leur incertitude adaptées au bâti français sont proposées. Par ailleurs nous avons comparé la partie linéaire des courbes de capacité utilisées dans la méthode européenne Risk-UE avec les fréquences mesurées. Il apparaît que la variabilité est très importante. Les courbes fournies ne sont, en outre, pas toujours en cohérence avec la réalité du bâti français. Les enregistrements de vibrations in situ peuvent donc notamment être un outil facile à mettre en œuvre pour calibrer la partie linéaire des courbes de capacité.*

ABSTRACT. *Large-scale seismic vulnerability assessment methods use simplified formulas and curves, often without providing uncertainties. They are seldom compared to experimental data. Therefore, we recorded ambient vibrations and estimated modal parameters (resonance frequencies, modal shapes and damping) of 60 buildings in Grenoble (France) of various types (masonry and reinforced concrete).*

*The knowledge of resonance frequencies in the linear domain is essential in the seismic design. Hence, we compared resonance frequency formulas given in the design code with this experimental data. The variability is underestimated and only two parameters (type and height of the building) seem to be statistically significant. Moreover, we compared the linear part of capacity curves used in European Risk-UE method to the measured frequencies. The variability is still very large and these curve are often not relevant for the French buildings. As a result, ambient vibration recordings may become an interesting tool in order to calibrate the linear part of capacity curves.*

MOTS-CLÉS : *vulnérabilité sismique à grande échelle, vibrations ambiantes, analyse modale, fréquences propres, courbes de capacité.*

**1. Introduction**

L'étude de la vulnérabilité sismique à grande échelle et le respect des règles parasismiques pour les bâtiments courants fait appel à des formules et des courbes dont les données sources et les méthodes de calculs appliquées ont rarement fait l'objet d'une publication. Ainsi, elles sont utilisées la plupart du temps sans précautions, parfois hors de leur domaine d'application, ce qui peut conduire à des interprétations erronées. La facilitation de l'acquisition et du traitement de données in situ dans les structures de génie civil permet aujourd'hui de disposer d'une quantité raisonnable de données réelles et fiables à comparer avec ces formules, obtenues soit par calcul, soit par mesures de laboratoire. Dans le présent article, nous nous sommes intéressés à la première fréquence de résonance dans les bâtiments courants. Dans les codes parasismiques de tous les pays, des formules simplifiées permettent de la calculer à moindres frais. Cette valeur de fréquence dans le « domaine des petites oscillations » (PS92, 1995) est fondamentale pour l'estimation du mouvement à prendre en compte dans les calculs de dimensionnement. Par ailleurs, dans les méthodes d'estimation de la vulnérabilité à grande échelle comme HAZUS (FEMA, 1999) aux Etats-Unis ou Risk-UE (Risk-UE, 2003) en Europe, chaque type de bâtiment est modélisé par une courbe de capacité qui relie les contraintes aux déformations. La partie élastique de cette courbe est théoriquement une droite de pente le carré de la pulsation propre du bâtiment. L'objectif de cette étude est donc de confronter les fréquences propres de bâtiments obtenues par des enregistrements in situ de vibrations ambiante aux données théoriques de la méthode Risk-UE. Il s'agit également d'interpeller les utilisateurs de ces courbes pour leur faire prendre conscience de leurs limitations.

**2. Fréquences de résonance obtenues par mesures in situ.**

Les structures de génie civil sont sollicitées en permanence par sol (bruit de fond sismique), l'atmosphère (vent) et les activités humaines internes (piétons, ascenseurs…). Le plus souvent, c'est le bruit de fond sismique, liés aux océans, à l'atmosphère et aux activités humaines, qui domine cette sollicitation. Il a la particularité d'être un bruit blanc, c'est-à-dire d'avoir un spectre plat. L'enregistrement des vibrations ambiantes au dernier étage d'une structure, où l'amplitude est la plus forte, permet de déterminer ses fréquences de résonance, caractéristiques intrinsèques de la structure liées à la répartition de ses masses et de ses rigidités. En petites déformations, cette fréquence est quasiment constante, avec une légère non-linéarité liée à l'ouverture élastique de fissures lorsque l'amplitude de la sollicitation augmente (Dunand et al., 2006, Michel et Guéguen, 2007). La fréquence obtenue sous bruit de fond est donc pertinente pour l'étude de la partie élastique du dimensionnement et de la vulnérabilité sismique d'une structure.

Les données que nous allons utiliser ici ont trois origines. Une première campagne de mesure a été réalisée en 1995 par Mohammed Farsi et Pierre-Yves Bard (Farsi, 1996, Farsi et Bard, 2004) au LGIT (Laboratoire de Géophysique Interne et Tectonophysique) de Grenoble. Ils ont enregistré 30 minutes de vibrations ambiantes dans 42 bâtiments de Grenoble, dont 39 en béton armé et 3 en maçonnerie, à leur sommet et leur base. Les stations utilisées sont des RefTek 6 composantes avec des capteurs Guralp CMG5 (réponse plate en accélération de 0.1 à 50 Hz) et Mark products L22 (réponse plate en vitesse de 2 à 50 Hz). La fréquence de résonance de chaque bâtiment dans ses directions principales a été obtenue par pointé des pics sur les fonctions de transfert. Ces dernières ont été calculées après une moyenne des transformées de Fourier sur des fenêtres de bruit stationnaire. Seuls les bâtiments en béton ont été utilisés (paragraphe 3 et 4).

En 2002, dans le cadre du projet GEMGEP, le CETE Méditerranée a enregistré des vibrations ambiantes au dernier étage et au rez-de-chaussée de 54 bâtiments de Nice, 25 en maçonnerie et 28 en béton armé. Ils ont utilisé des stations Lennartz et Hathor et des capteurs Lennartz 3D-5s pour des durées d'enregistrement de 15 minutes. Les premières fréquences propres ont été déterminées à l'aide de la méthode du décrément aléatoire (Dunand, 2005). Seuls les bâtiments en béton (paragraphe 3) ont été utilisés dans la présente étude.

Enfin, nous avons mené trois campagnes de mesures en 2004, 2005 et 2006 dans des bâtiments grenoblois. L'utilisation d'une station portable Cityshark II (Châtelain et al., 2000) permettant l'acquisition de 18 voies (6 capteurs vélocimétriques Lennartz 3D-5s, enregistrements de 15 min) a conduit à une étude plus complète sur chaque bâtiment, avec au moins un enregistrement par étage. Au total, 60 bâtiments ont été étudiés 33 en maçonnerie et 27 en béton armé. Parallèlement, les caractéristiques des structures ont été relevées dans le but d'appliquer une étude de vulnérabilité à grande échelle (Guéguen et Michel, 2007). Les fréquences propres ont été calculées par la méthode de la Frequency Domain Decomposition (FDD, Brincker, 2001) qui permet une détermination plus précise et robuste des modes de vibration, même s'ils sont proches.

## 3. Formules analytiques de fréquence de résonance pour les bâtiments en béton

### 3.1. *Formules existantes dans les codes parasismiques*

Dans les règles parasismiques, la période propre du bâtiment est nécessaire pour l'évaluation de la sollicitation à prendre en compte. Il s'agit de la fréquence dans le domaine des petites déformations comme cela est rappelé dans les règles américaines et PS92, donc équivalente à celle obtenue sous vibrations ambiantes. En effet, les non linéarités qui affectent les fréquences propres des bâtiments dans le domaine élastique sont faibles (Hans et al., 2005 ; Michel et Guéguen, 2007). Pour aider le concepteur, des formules empiriques donnant la période propre sont précisées dans les codes. Farsi et Bard (2004) ont déjà comparé leurs mesures aux règles américaines (UBC88), françaises (PS92) et algériennes (RPA88). La formule la plus simple utilisée généralement est T=N/10, avec N le nombre d'étages.

Les formules américaines sont initialement issues de la campagne d'enregistrements de vibrations ambiantes de Carder (1936) après le séisme de Long Beach de 1933 sur 212 bâtiments, complétée avec des mesures japonaises (Housner, 1963). Cela concernait des bâtiments en maçonnerie et en acier essentiellement. Les formules qui en ont découlé (T ∝ H/√L, avec H la hauteur et L la longueur du bâtiment parallèlement à la direction de sollicitation considérée) ont été largement débattues dans les années 60 (Housner, 1963). Ce type de formule a également été adopté dans le code algérien (RPA88), coréen (Lee et al., 2000) et partiellement dans les règles françaises PS92, où elle est corrigée d'un facteur lié à l'élancement pour prendre en compte la raideur supérieure des murs voiles ou des remplissages. Les formules des PS92 sont :

$$T = 0.10 \frac{H}{\sqrt{L}} \text{ (ossatures non remplies)} \qquad [1]$$

$$T = 0.08 \frac{H}{\sqrt{L}} \sqrt{\frac{H}{H+L}} = 0.08 \frac{H}{\sqrt{L}} \sqrt{\frac{1}{1+\frac{L}{H}}} \text{ (murs voiles ou mixtes)} \qquad [2]$$

$$T = 0.06 \frac{H}{\sqrt{L}} \sqrt{\frac{H}{H+2L}} = 0.06 \frac{H}{\sqrt{L}} \sqrt{\frac{1}{1+2\frac{L}{H}}} \text{ (ossatures remplies)} \qquad [3]$$

Housner (1963) et Farsi et Bard (2004) critiquent la pertinence de l'utilisation de la dimension latérale L dans la formule. Pour ces auteurs, seule la hauteur du bâtiment a une influence statistique sur sa période propre, si l'on veut se contenter de formules simplifiées. Bien que Carder ait été un précurseur dans les mesures in situ, les techniques d'enregistrement et de détermination des fréquences propres en étaient à leurs balbutiements (enregistrements analogiques, détermination de la période en comptant les oscillations…) et le bâti a lui même largement évolué. Pourtant aucune autre campagne de cette ampleur n'a été menée depuis et ces données ont été utilisées pendant longtemps. Les conclusions de Housner ainsi que les enregistrements du séisme de San Fernando en 1971 dans des structures grâce au programme californien CSMIP ont fait évoluer les formules dans les versions suivantes de l'Uniform Building Code. Elle prend la forme suivante :

$$T = C_t H^\beta \qquad [4]$$

Ce type de formule est également utilisé dans l'Eurocode 8. Différents coefficients $C_t$ et $\beta$ sont donnés selon les codes et les types de bâtiments. L'UBC 97 et l'Eurocode 8 utilisent $\beta=0.75$, l'Eurocode limitant l'emploi de cette formule aux bâtiments de hauteur supérieure à 40m et proposant $C_t=0.05$ pour les voiles. Le document FEMA356 (FEMA, 2000) propose $\beta=0.9$ pour les ossatures en béton armé, $\beta=0.75$ pour les autres types de contreventement (ossatures en acier mis à part). Dans le cadre du projet Risk-UE, Lagomarsino et Giovinazzi (2006) ont utilisé pour déterminer la partie linéaire de la courbe de capacité des bâtiments en béton armé $C_t=0.065$ et $\beta=0.9$.

Dans l'esprit des codes américains, cette formule est censée sous-estimer les périodes de 10 à 20% de manière à calculer des valeurs d'effort tranchant conservatives (Goel et Chopra, 1998). Notons que l'on sous-estime alors les déplacements, ce qui n'est pas forcément un bon calcul compte tenu des nouvelles approches du génie parasismique, fondées sur les méthodes en déplacement. Or Goel et Chopra (1998) critiquent cette formule du fait qu'elle ne sous-estime pas les périodes, d'une part, et qu'elle n'est pas satisfaisante pour les structures en voiles. Ils proposent la formule suivante, plus complexe et dépendant des dimensions des murs :

$$T = C_t \frac{H}{\sqrt{\overline{A}_e}} \text{ avec } \overline{A}_e = \frac{100}{L \times l} \sum_{i=1}^{Nm} \left(\frac{H}{H_i}\right)^2 \frac{A_i}{1+0.83\left(\frac{H_i}{L_i}\right)^2} \qquad [5]$$

avec $A_i$, $H_i$ et $L_i$ respectivement l'aire en plan, la hauteur et la longueur du mur i, l est la largeur du bâtiment. Cette méthode est obtenue par calculs théoriques et validée sur 17 bâtiments californiens d'après les enregistrements de séismes de 1971 (San Fernando) à 1994 (Northridge). $\overline{A}_e$ représente la surface de cisaillement équivalente, en pourcentage de la surface en plan du bâtiment. Cette méthode nécessite cependant d'étudier plus en détail le bâtiment, en particulier d'avoir accès à ses plans, ce qui est le cas pour un bâtiment à dimensionner mais rarement pour un bâtiment existant.

Depuis les travaux précurseurs de Carder, toutes les formules proposées ont été validées sur un très petit nombre de structures réelles (typiquement une vingtaine). Une mise en commun des bases de données de périodes mesurées en structure serait une aide considérable pour de nombreux travaux en génie parasismique.

### 3.2. *Régression linéaire multiple sur les données*

Compte tenu des formules précédentes, trois points nous ont semblé importants à clarifier. D'une part, le paramètre L a-t-il sa place dans une formule donnant la période ? D'autre part, quel exposant faut-il affecter au paramètre H puisque certaines formules donnent 1 ou 0.9, d'autres ¾ ? Enfin, le nombre d'étages, plus facile à obtenir, est-il aussi pertinent que la hauteur pour l'estimation des périodes dans une formule simplifiée ? Un test sur les formules plus complexe comme celle de Goel et Chopra (1998) n'a pas été possible car les plans des bâtiments n'étaient pas accessibles. Le système structural de tous les bâtiments n'était également pas disponible, cependant, la très grande majorité des structures étudiées sont composées de murs voiles et les essais visant à discriminer des bâtiments ayant une ossature remplie n'ont pas amené de résultats pertinents. Les résultats présentés ici concernent donc l'intégralité des bâtiments en béton de Grenoble et Nice qui sont majoritairement des structures en murs voiles, soit 173 périodes propres.

Nous avons choisi d'étudier le logarithme des périodes en fonction des logarithmes des paramètres hauteur, nombre d'étages et longueur de la façade parallèlement à la direction considérée soit H, N, et L respectivement. Il faut noter que le paramètre pertinent n'est pas N, mais N+1 car c'est le nombre de niveaux qui est équivalent à la hauteur. Nous avons donc réalisé une régression linéaire multiple du logarithme de la période en fonction de ces paramètres. On calcule la matrice de corrélation des données (c'est-à-dire la matrice de covariance des données centrées réduites) qu'il suffit d'inverser. Les coefficients de la régression linéaire ainsi que les coefficients de corrélations (total et partiels) sont ensuite calculés en fonction des éléments de cette matrice inversée.

Les régressions en fonction de H et L et N+1 et L ont donné à L un coefficient de corrélation partielle avec les données de 13%, ce qui signifie que L est très peu corrélé aux périodes. Par ailleurs, le coefficient de corrélation totale n'est pas amélioré en prenant ce paramètre en compte. Les coefficients de la régression donneraient alors un exposant de 0.05, ce qui n'est pas du tout en accord avec le -0.5 trouvé dans les formules précédemment décrites. On peut en conclure, comme de nombreux auteurs, que le paramètre L n'est pas pertinent d'un point de vue statistique dans l'explication de la variance des périodes propres.

La hauteur et le nombre de niveaux expliquent donc, seuls, entre 85% et 90% de la variance des périodes. La différence entre une formule utilisant la hauteur et une formule utilisant le nombre de niveaux est de quelques pourcents, la hauteur restant le paramètre le plus pertinent. Dans le cadre d'une formule simplifiée, ces deux paramètres ont donc une qualité sensiblement égale.

Les exposants liés à la hauteur H et au nombre de niveaux ont été calculés respectivement à 0.98 et 0.92. Il semble donc clair que c'est un exposant de 1 qui doit être affecté à la hauteur ou au nombre de niveaux, comme préconisé par Goel et Chopra (1998), le 0.9 du document FEMA356 convenant également, et non 0.75 comme dans l'Eurocode et l'UBC97. On revient donc à une régression linéaire à un paramètre avec l'hypothèse que l'ordonnée à l'origine est nulle :

$$T = \frac{H}{74.3} = \frac{N+1}{25.5} \quad \text{pour les murs voiles, } \sigma^* = \{0.08\,; 0.09\} \text{ (en s.)} \qquad [6]$$

Cette estimation, avec l'intervalle de confiance à 95% pour chaque hauteur (Figure 1), montre que l'Eurocode 8 ($1/C_t=20$, valeur pour les voiles, contre 74.3 trouvés ici) n'est pas en accord avec la réalité du bâti français (y compris pour H>40m) et surestime les périodes. La formule de Lagomarsino et Giovinazzi (2006), dans le cadre du projet Risk-UE, a pour objectif de surestimer les périodes pour être conservatif vis-à-vis des déplacements, mais la comparaison avec les données réelles ($1/C_t=15$) montre que cette surestimation est importante en ce qui concerne les bâtiments français ce qui conduit à estimer des dommages bien supérieurs à la réalité.

Les formules comme celle de l'Eurocode 8 ne sont pas fournies avec un écart type, il n'est donc pas possible de tester la plausibilité de l'appartenance de notre jeu de données à leur modèle statistique. C'est pourquoi, nous proposons avec nos formules un écart type estimé comme suit :

$$\sigma^* = \sqrt{\frac{1}{n-2} \sum_{i=1}^{n} \left(T_i - T_i^{est}\right)^2} \quad \text{avec } T_i^{est} \text{ la période estimée par le modèle} \qquad [7]$$

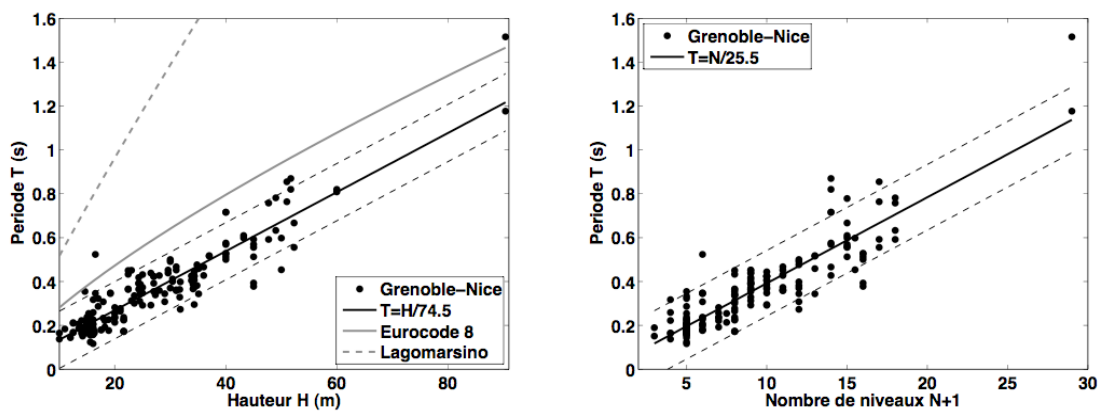

**Figure 1.** *Premières périodes propres des bâtiments de Grenoble et Nice (points noirs) en fonction de la hauteur de la structure (à gauche) et de son nombre de niveaux (à droite), leur régression linéaire (droites noires) et l'intervalle de confiance à 95% pour chaque abscisse (droites pointillées noires). En gris, en trait plein la formule de l'Eurocode 8 avec $C_t=0.05$ et en trait pointillé la formule utilisées par Lagomarsino et Giovinazzi (2006) dans Risk-UE.*

## 4. Partie élastique des courbes de capacité dans les méthodes d'analyse de vulnérabilité à grande échelle

### 4.1. *Méthodes d'analyse de vulnérabilité à grande échelle*

Dans les méthodes d'analyse de vulnérabilité à grande échelle, la donnée de base n'est pas le bâtiment mais le type. La première étape dans la mise au point d'une telle méthode est donc de définir une typologie représentative du lieu que l'on souhaite étudier. Dans le cas des méthodes HAZUS (FEMA, 1999) aux Etats-Unis et Risk-UE (2003) en Europe, les bâtiments sont séparés en fonction de leur système structural. Les typologies sont donc simples, mais recouvrent de nombreux cas différents liés à la qualité des matériaux utilisés, du savoir faire de chaque pays voire de chaque région.

Chaque type est ensuite modélisé à l'aide d'une courbe de capacité qui relie la force appliquée aux déplacements de la structure. Pour Risk-UE, comme pour HAZUS, ces courbes sont bilinéaires : la première droite représente la partie élastique jusqu'au point de plastification (yield point), la deuxième, de pente plus faible, représente le comportement ductile jusqu'à la ruine. Trois courbes sont données pour chaque typologie selon la hauteur des structures : les structures basses (jusque 2 étages pour la maçonnerie, 3 pour le béton), les structures moyennes (3 à 5 étages pour la maçonnerie, 4 à 7 pour le béton) et les structures hautes (6 étages et plus pour la maçonnerie, 8 et plus pour le béton).

La courbe de capacité permet de déterminer directement le point de performance du bâtiment pour un mouvement sismique donné, représenté par son spectre de réponse dans un plan (Sd,Sa) et donc d'en déduire l'état de dommage attendu (Risk-UE, 2003).

### 4.2. *Comparaison avec les données in situ*

Pour un oscillateur linéaire à 1 degré de liberté, la force F s'écrit en fonction du déplacement U :

$$F = KU = \omega^2 MU \qquad [10]$$

avec K la rigidité, M la masse et ω la pulsation propre. Les courbes de capacité sont représentées en accélération A, soit :

$$A = \frac{F}{M} = \omega^2 U \qquad [11]$$

La première droite qui représente la partie élastique a donc pour pente $\omega^2=4\pi^2f^2$ avec f la fréquence propre. On peut donc comparer les droites théoriques avec des droites issues des fréquences mesurées. Nous avons choisi d'arrêter la partie élastique au déplacement Dy donné par Risk-UE. En pratique, comme la courbe de capacité est bilinéaire, elle sous-estime légèrement la fréquence aux petites oscillations pour être plus pertinente aux alentours du point de plastification. On s'attend donc à obtenir des courbes théoriques un peu en dessous des mesures.

4.2.1. *Bâtiments en béton*

Parmi les bâtiments de Grenoble étudiés, deux classes de la typologie Risk-UE sont présentes : le type murs (RC2) et le type ossature avec remplissage en maçonnerie (RC3.1). Nous avons comparé aux courbes données par Lagomarsino et Giovinazzi (2006) les courbes de capacité issues des fréquences mesurées. Les résultats sont présentés sur la figure 2. On peut remarquer tout d'abord que les données, correspondant à une seule courbe théorique, sont très dispersées. Aucune incertitude n'est pourtant fournie avec ces courbes de capacité. Pourtant, entre deux bâtiments de 4 et de 7 étages du même type, la fréquence peut changer du simple au double (Figure 1), sans parler de variation dans les dispositions constructives. La pente peut alors varier du simple au quadruple et il en est de même pour le point de plastification.

Pour le type RC2, la quantité de donnée est suffisante. Comme prévu les courbes théoriques sont situées dans la fourchette basse des valeurs *in situ* (Figure 2). Les estimations du point de plastification sont donc bien conservatives, mais elle vont jusqu'à diviser par 5 l'accélération que le bâtiment peut supporter élastiquement pour un seuil en déplacement donné.

Pour le type RC3.1, la quantité de donnée est assez faible, mais les conclusions sont similaires au type RC2. Il y a peu de différence de comportement entre les deux types, aussi bien en ce qui concerne les données expérimentales que les courbes Risk-UE. Il faut cependant noter que les courbes fournies par l'Université de Thessalonique dans le document Risk-UE WP4 (2003) sont très différentes pour ce type et semblent moins bien représenter les données que les courbes de Lagomarsino et Giovinazzi (2006). Les types Risk-UE peuvent donc regrouper des bâtiments avec des dispositions constructives assez différentes d'un pays à l'autre.

4.2.2. *Bâtiments en maçonnerie*

Le type M1.1 Risk-UE (pierre tout venant) est identique dans sa description au type MA1 de la Base de Données Typologique de Grenoble (Guéguen et Vassail, 2004). Nous avons regroupé dans le type M1.2 Risk-UE (pierre brute) les types MA2 à MA5 et MA8 de la BDT qui détaillent les évolutions de l'utilisation de la pierre brute à Grenoble. Elle regroupe ainsi la plus grande partie des bâtiments étudiés. Il est également intéressant d'étudier le type M3.4 (maçonnerie avec planchers en béton), identique au type MA7. Les courbes de capacité Risk-UE ont été proposées par l'Université de Gênes par Lagomarsino et Giovinazzi (2006), selon une méthode indépendante d'une formule donnant la période en fonction de la hauteur.

Les bâtiments en pierre tout venant (M1.1) sont supposés plus souples que les bâtiments en pierre brute (M1.2) selon Risk-UE, ce qui ne semble pas être le cas pour le bâti grenoblois (Figure 2). La courbe Risk-UE du type M1.2 (pierre brute) est très proche de la moyenne des courbes expérimentales. Elle représente donc bien le bâti, mais, de fait, n'est plus conservative vis-à-vis des dommages. Enfin, il est intéressant de remarquer que la courbe théorique représentant le type M3.4 (maçonnerie plancher béton) suppose une structure plus souple (ou de masse plus importante) alors que les mesures tendent à montrer qu'elle est aussi rigide, voire plus rigide que le type M1.2.

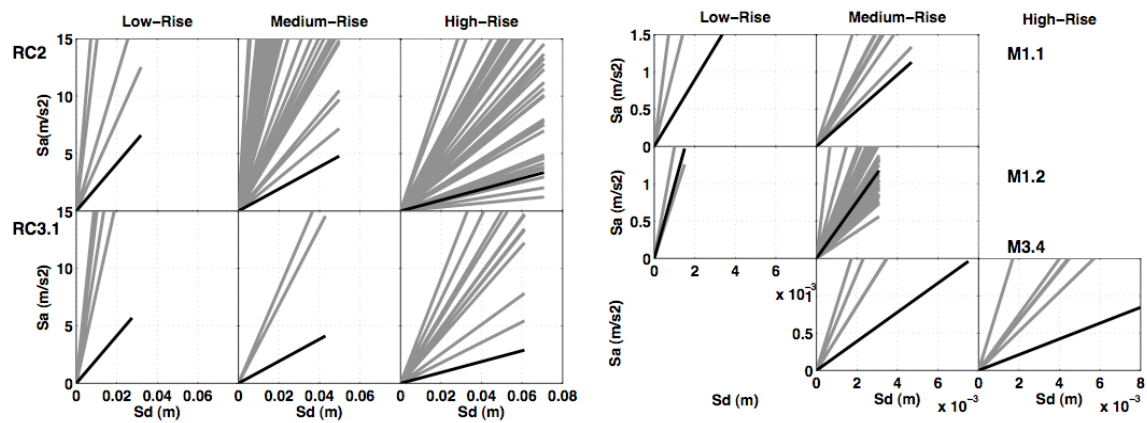

**Figure 2.** *Partie linéaire des courbes de capacité expérimentales (gris) comparées aux courbes Risk-UE de Lagomarsino et Giovinazzi (2006) (noir). À gauche, bâtiments en béton (type RC2=murs voiles, type RC3.1=ossatures avec remplissage), à droite, bâtiments en maçonnerie (type M1.1=pierre tout venant, type M1.2=pierre brute, type M3.4=pierre brute avec planchers en béton)*

## 5. Conclusion

L'utilisation de 173 périodes propres de bâtiments en béton et 66 périodes de bâtiments en maçonnerie, mesurées à Grenoble et Nice a permis de mettre en évidence la variabilité de ce paramètre essentiel en génie parasismique. La fréquence en petites déformations permet en effet de calculer l'état élastique de la structure pour un spectre de réponse donné. Lors du dimensionnement d'une structure en béton, la formule proposée par l'Eurocode 8, issue directement des règles américaines, surestime largement les périodes réelles. Elle est donc conservative vis-à-vis des déplacements, ce qui va dans le sens de la sécurité si on utilise une méthode en déplacement, mais sous-estime les forces. Or bien que la tendance s'inverse, la plupart des praticiens utilisent encore des méthodes en force. Nous proposons les formules simplifiées $T=H/74.3=(N+1)/25.5$ avec comme écarts types respectifs $\sigma^*$ 0.08 et 0.09 s. Le nombre de niveaux ($N+1$) et la hauteur $H$ sont aussi bien corrélés à la période propre $T$.

Les courbes de capacité de la méthode Risk-UE sont directement liées à la période propre. Une grande partie de l'incertitude sur celles-ci pourrait être évitée pour les bâtiments en béton en donnant une courbe par nombre d'étages à l'aide des régressions proposées. La typologie utilisée pour une méthode n'est pas forcément transposable telle qu'elle pour une autre région. Ainsi, la maçonnerie de Grenoble est plus homogène que les types proposés dans Risk-UE. La sous-estimation volontaire des périodes dans la méthode Risk-UE induit nécessairement une surestimation des dommages. Cette surestimation est nécessaire par souci de sécurité, mais perd de son sens si l'on n'inclut pas une estimation de la variabilité du modèle.

Ces résultats plaident donc en faveur de la mise en commun au niveau européen des données acquises par les différents groupes travaillant sur la vulnérabilité des structures, que ce soit des données structurales comme des mesures de périodes propres ou d'autres tests mécaniques. Les enregistrements restent cependant la seule méthode permettant d'estimer les périodes propres des structures avec précision. Ils devraient donc être un outil privilégié pour la calibration de la partie linéaire des courbes de capacité.

## 6. Bibliographie